\documentclass{aa}   
\usepackage{txfonts}	
\usepackage{graphicx}			
\usepackage{natbib}	

\begin{document}

\title{The protoMIRAX Hard X-ray Imaging Balloon Experiment}
\authorrunning{J. Braga et al.}
\titlerunning{The protoMIRAX balloon experiment}
\author{Jo\~ao Braga, Flavio D'Amico, {Manuel A.\ C.\ Avila}, Ana V. Penacchioni, {J.\ Rodrigo Sacahui}, Valdivino A.\ de Santiago Jr., F\'atima Mattiello-Francisco, Cesar Strauss, M\'arcio A.\ A.\ Fialho}
\institute{National Institute of Space Research -- INPE, Av. dos Astronautas 1758, S\~ao Jos\'e dos Campos, SP, Brazil, 12227--010}
\offprints{\email{joao.braga@inpe.br}}

\abstract{The protoMIRAX hard X-ray imaging telescope is a balloon-borne experiment developed as a pathfinder for the MIRAX satellite mission. The experiment consists essentially in a coded-aperture hard X-ray (30-200 keV) imager with a square array (13$\times$13) of 2mm-thick planar CZT detectors with a total area of 169\thinspace cm$^2$. The total, fully-coded field-of-view is 21$^{\circ}\times 21^{\circ}$\ and the angular resolution {is} 1$^{\circ}$43\arcmin. In this paper we describe the protoMIRAX instrument and all the subsystems of its balloon gondola, and we show simulated results of the instrument performance.
}
{The main objective of protoMIRAX is to carry out imaging spectroscopy of selected bright sources to demonstrate the performance of a prototype of the MIRAX hard X-ray imager. 
}
{Detailed background and imaging simulations have been performed for protoMIRAX balloon flights. The  $3$-$\sigma$ sensitivity for the 30-200 keV range is $\sim $1.9 $\times$ 10$^{-5}$ photons cm$^{-2}$ s$^{-1}$ for an integration time of 8\thinspace hs at an atmospheric depth of 2.7 g cm$^{-2}$ and an average zenith angle of 30$^{\circ}$. We have developed an attitude control system for the balloon gondola and new data handling and ground systems that also include prototypes for the MIRAX satellite. 
}
{We present the results of Monte Carlo simulations of the camera response at balloon altitudes, showing the expected background level and the detailed sensitivity of protoMIRAX. We also present the results of imaging simulations of the Crab region.
}
{The results show that protoMIRAX is capable of making spectral and imaging observations of bright hard X-ray source fields. Furthermore, the balloon observations will carry out very important tests and demonstrations of MIRAX hardware and software in a near space environment.}


\maketitle

\section{Introduction}

Coded mask experiments have been widely used in astronomical hard X-ray ($E \gtrsim 10$\thinspace keV) observations of the sky due to several factors. First, the technology to develop focussing telescopes above these energies has only recently been mastered and implemented, especially in the NuSTAR observatory \citep{2013ApJ...770..103H}. {The ASTRO-H mission, expected to be launched in the near future, will also make use of focussing optics in the hard X-ray range through its Hard X-ray Telescope (HXT) \citep{2014SPIE.9144E..25T}.}  While providing much higher sensitivities and better angular resolution in general, these instruments require very long focal lengths and sophisticated control and alignment systems. In addition, the highest energy achievable is still significantly below 100 keV, {even though there are promising higher-energy ($\lesssim$\thinspace 600\thinspace keV) focussing techniques like Laues lenses being developed \citep{2013SPIE.8861E..07V} .} Coded mask instruments, on the other hand, allow imaging over very wide fields-of-view (which can be a significant fraction of the sky) and up to energies of several hundred keV. Another factor is that coded mask instruments can be compact and easy to implement, making them good options for wide-field hard X-ray and low energy $\gamma$-ray monitors of the highly variable and transient source populations in this energy range. Important examples of coded mask satellite instruments that have made significant contributions to astronomy are: SIGMA \citep{1991AdSpR..11..289P} and ART-P \citep{1990AdSpR..10..233S} instruments onboard GRANAT, the ASM on RXTE \citep{1996ApJ...469L..33L}, the WFCs on BeppoSAX \citep{1997A&AS..125..557J}, {the WFM on HETE-2} \citep{2003AIPC..662....3R}, instruments on the INTEGRAL satellite \citep{1995ExA.....6...71W}, and BAT/SWIFT \citep{2004NewAR..48..431G}.

In this paper we describe the protoMIRAX balloon experiment, a wide field hard X-ray coded mask imager under development at the National Institute for Space Research (INPE) as a pathfinder for the MIRAX (Monitor e Imageador de Raios X) satellite mission \citep[see also AIP Conference Proceedings 840, 2006, for a series of papers and information about MIRAX]{2004AdSpR..34.2657B}. The main instrument of the balloon payload is a hard X-ray camera that employs an array of 13$\times$13 CdZnTe (``CZT'') {detectors} as the position-sensitive detector plane. The experiment will be launched on a stratospheric balloon over Brazil at a latitude of $\sim23^{\circ}$\thinspace S to carry out imaging spectroscopy of selected bright sources. In addition to testing the camera perfomance in a near space environment, the experiment is a testbed for new space technologies being developed at INPE. A new attitude control and pointing system for scientific balloon gondolas, including two new star trackers, will be tested, as well as novel data handling and ground support systems.

In section \ref{sec:science} we present the scientific objectives of the experiment. In section \ref{sec:experiment} we describe the X-ray camera and the other subsystems that protoMIRAX comprises. In section \ref{sec:bkg} we present the results of Monte Carlo simulations of the camera response at balloon altitudes, showing the expected background level and the detailed sensitivity of protoMIRAX. A simulated image of the Crab region is shown in section \ref{sec:image}. We present our conclusions in section \ref{sec:conclusion}. 

\section{Scientific Objectives} 
\label{sec:science}

The protoMIRAX experiment will perform imaging spectroscopy with good energy resolution of bright Galactic hard X-ray sources such as the Crab Nebula (and its pulsar) and the Galactic Center complex. 

In X-ray astronomy, the Crab has been considered as a standard candle due to its large and nearly constant flux at Earth {(see {\citet{2014RPPh...77f6901B} and \citet{ 2015ApJ...801...66M} for recent reviews)}}. At X-ray energies above $\sim$30{\thinspace}keV, the Crab is generally the strongest persistent source in the sky and has a diameter of $\sim$1{\thinspace}arcmin. The Crab has experienced some recent flares in the GeV range \citep{2011Sci...331..736T} which were simultaneous or just previous to a particularly active period in hard X-rays \citep{2011ApJ...727L..40W}. Those surprising variations are of the order a few percent on the 50-100 keV energy range. ProtoMIRAX will be able to measure the Crab spectrum from 30 to 200 keV. These observations will also be used for flux calibrations and imaging demonstrations. 
 
The Galactic Center (GC) region is very favorable for protoMIRAX observations due to its declination of $\delta \sim -29^{\circ}$, since the balloon will fly over a region at latitudes of $\sim -23^{\circ}$. One of the most interesting sources is 1E{\thinspace}1740.7$-$2942, the brightest and hardest persistent X-ray source within a few degrees of the GC. Due to a similar hard X-ray spectrum and comparable luminosity to Cygnus X-1 \citep{1984SSRv...38..353L}, 1E{\thinspace}1740.7$-$2942\ is classified as a black hole candidate. Because of the two-sided radio jets associated to the source, the object was dubbed the first ``microquasar" \citep{1992Natur.358..215M}, an X-ray binary whose behavior mimics quasars on a much smaller scale. Since the GC direction has extremely high extinction, a counterpart has not yet been identified despite deep searches made in both the optical and IR bands. The source was recently studied by our group from soft to hard X-rays up to 200{\thinspace}keV \citep{2014A&A...569A..82C}, showing spectra that can be very well modeled by thermal Comptonization of soft X-ray photons. We will image the Galactic Centre region with protoMIRAX with 1E{\thinspace}1740.7$-$2942\ in the centre of the field-of-view. {Our observations in the hard X-ray band could be useful to measure flux and spectral parameters}, especially because the source is most of the time in the hard state.

Another interesting source that will be in the same field-of-view is GRS{\thinspace}1758$-$258, also a microquasar, and one of the brightest X-ray sources near the GC at energies greater than $50$ keV \citep{1991ApJ...383L..49S}. Like its sister source 1E{\thinspace}1740.7$-$2942, GRS{\thinspace}1758$-$258\ spends most of the time in the hard state. Its hard X-ray spectra and variability are also similar to that of Cyg{\thinspace}X-1 \citep{1996rftu.proc..157K}. GRS{\thinspace}1758$-$258\ also does not have yet an identified counterpart, so its mass and orbital periods are still unknown. {The protoMIRAX observations will in principle be capable to attain spectral and flux information} about GRS{\thinspace}1758$-$258.

The GC field to be observed by protoMIRAX will also include GX{\thinspace}1$+$4, the best studied accreting pulsar around the GC, with a pulse period of the order of 2 minutes. This object is the prototype of the small but growing subclass of slow-rotating accreting X-ray pulsars called Symbiotic X-ray Binaries (SyXB), in which a neutron star accretes from the wind of a M-type giant companion \citep{2006A&A...453..295M}. GX{\thinspace}1$+$4\ is a persistent source, but with strong, irregular flux variations on various timescales and extended high/low states. Its peculiar spin history has been the object of intensive study \citep{1989PASJ...41....1N}, and \citet{1988Natur.333..746M} have shown a clear transition from spin-up to spin-down behaviour at the equilibrium period, characterizing a torque reversal episode. More recent studies \citep{2012A&A...537A..66G} have used data from {\it Beppo}SAX, INTEGRAL, {\it Fermi\/} and {\it Swift}/BAT to show that the source continues its spin-down trend with a constant change in frequency, and the pulse period has increased by $\sim$50\% over the last three decades. {Since the source is highly variable, our spectral and timing observations can possibly contribute to the flux and period histories for this peculiar object.}

Besides providing new data on astrophysical sources, the experiment will be able to make precise measurements of the hard X-ray background at balloon altitudes for the Brazilian mid-latitudes, which is important for investigations concerning the South Atlantic Anomaly (SAA) and Space Weather studies. 

\section{The protoMIRAX experiment}
\label{sec:experiment}

The protoMIRAX experiment consists basically in an hard X-ray imaging camera mounted in a stratospheric balloon gondola capable of carrying out pointing observations of selected target regions. Table \ref{tab:overview} presents an overview of the experiment baseline parameters.

\begin{table*}[!ht]
\centering
\caption{protoMIRAX baseline parameters}
\label{tab:overview}
\begin{tabular}{|l|l|}
\multicolumn{2}{c}{\textbf{Detector System}}\\
\hline
Detector type:  CdZnTe (CZT)  & Dimensions:  10\thinspace mm $\times$ 10\thinspace mm \\
\hspace{0.3cm} with platinum planar contacts & \hspace{0.3cm} $\times$ 2\thinspace mm (thickness)\\
Number of detectors:  169 ($13 \times 13$) & Gap between detectors:  $10$ mm\\
Energy Range:  $30 - 200$ keV & Geometrical area:  $169$ cm$^2$\\
Effective area @ 80 keV:  $52$ cm$^2$  & Time resolution:  5 $\mu$s\\
\hspace{0.3cm} {(through mask, at $2.7$ g cm$^{-2}$)} & \\
\hline
\multicolumn{2}{c}{\textbf{Collimator}} \\
\hline
Blade material:  Cu (0.5\thinspace mm),  & Cell size:  squares of 20-mm sides\\
\hspace{0.3cm} Pb (0.5\thinspace mm), Cu (0.5\thinspace mm) & External envelope:  260\thinspace mm \\
Blade height:  81\thinspace mm  & \hspace{0.3cm} $\times$ 260\thinspace mm $\times$ 81\thinspace mm\\
\hline
\multicolumn{2}{c}{\textbf{Coded Mask}} \\
\hline
Material:  1\thinspace mm-thick lead & Open fraction:  $0.497$\\
Basic pattern:  $13 \times 13$ MURA & Element Size:  $20$\thinspace mm $\times\ 20$\thinspace mm\\
Extended pattern:  $2 \times 2$ basic & Total mask dimensions:  500\thinspace mm $\times$ \\
\hspace{0.3cm} (minus 1 line and 1 column) & \hspace{0.3cm} 500\thinspace mm $\times$ 3\thinspace mm (total thickness)\\
Position: $650$\thinspace mm from detector plane & \hspace{0.3cm} (includes 2mm-thick acrylic substrate)\\
\hline
\multicolumn{2}{c}{\textbf{Shielding}} \\
\hline
Material:  lead (external -- 1.5\thinspace mm),  & Position:  around detectors and \\
\hspace{0.3cm} copper (internal -- 0.5\thinspace mm) & \hspace{0.3cm} collimator (sides and bottom)\\
\hline
\multicolumn{2}{c}{\textbf{Imaging Parameters}}\\
\hline
Angular Resolution:  $1^{\circ}43'$ & Total (fully-coded) FOV:  $21^{\circ} \times 21^{\circ}$ \\
Source Location Accuracy: 10$^{\prime}$ (10\thinspace $\sigma$) & \hspace{0.3cm} ($14.1^{\circ} \times 14.1^{\circ}$ FWHM)\\
\hline
\multicolumn{2}{c}{\textbf{Balloon gondola and flight}}\\
\hline
Gondola total mass:  $\sim 600$ kg & Gondola dimensions: 1.4\thinspace m $\times$ 1.4\thinspace m \\
Flight altitude:  $42$ km  (2.7 g cm$^{-2}$) &\hspace{0.3cm} $\times$ 1.8\thinspace m (height)\\
\hspace{0.3cm} at $-23^{\circ}$ latitude over Brazil & Single flight duration:  $\lesssim 40$\thinspace hours\\
Pointing accuracy:  $\sim$10 arcminutes & Star tracker precision:  10 arcseconds\\ 
\hline
\end{tabular}
\end{table*}

\subsection{The X-Ray Camera}

The X-ray camera (XRC) is a coded-aperture wide-field hard X-ray imager that consists in an aluminum tower-like structure that supports an array of X-ray detectors, a collimator, a coded mask, passive shields and other electrical shielding and supporting components. In this section we describe in details each component of the XRC.

\subsubsection{The X-ray detectors}

The X-ray detectors we use in this experiment are band-gap, room-temperature semiconductors made of an alloy of 90\% Cadmium Telluride (CdTe) and 10\% Zinc Telluride (ZnTe), called CdZnTe or simply CZT. They have high photoelectric efficiency up to hundreds of keV, due to the high atomic numbers of Cd ($Z=48$) and Te ($Z=52$) and a density of 6 g cm$^{-3}$. The probability of photoelectric absorption per unit pathlength is roughly a factor of 4 to 5 times higher than in Ge for typical gamma-ray energies. The relatively large band gap of 1.5\thinspace eV precludes significant thermal excitations at room temperature and provides good enough energy resolution ($\Delta E/E \lesssim 10$\% at 60\thinspace keV). 

CZT detectors have been often used in X-ray astronomy due to their high efficiency with small thickness (thus reducing background, which scales with volume) and relative ease of handling and mounting, allowing for tiling so that large nearly-contiguous detector planes can be built. 

The experiment uses 169 CZT detectors with platinum planar contacts and dimensions of 10\thinspace mm $\times$ 10\thinspace mm, thickness of 2\thinspace mm each, out of 200 units acquired from eV Products, USA. They will be configured in a $13 \times 13$ array, thus the total area of the detector plane is 169 cm$^2$. Due to physical mounting restrictions, the edges of adjacent detectors will be separated by 10\thinspace mm. The operational energy range will be from 30 to 200 keV, the lower limit determined by residual atmospheric absorption at balloon altitudes ($\sim$42\thinspace km) and the higher limit determined by detector photoelectric efficiency given its thickness. 

\begin{figure}[!ht]
\centering
\includegraphics[width=0.6\hsize,angle=0]{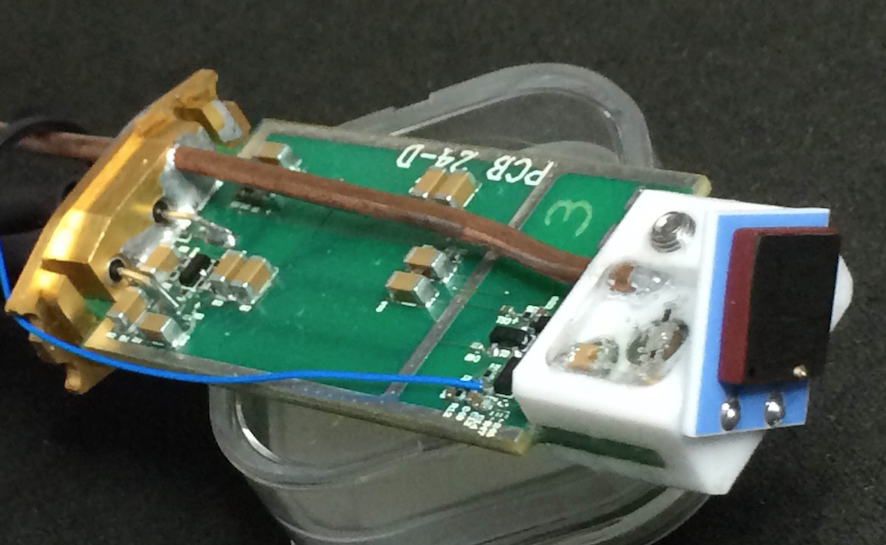}
\hspace{1cm}
\includegraphics[width=0.7\hsize,angle=0]{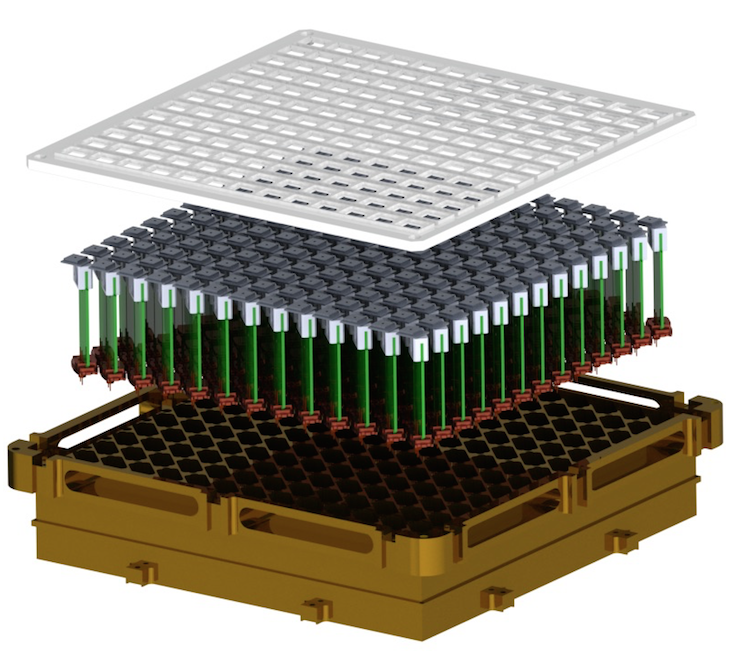}
\caption{{\it Top:\/} one CZT detector (20\thinspace mm$\times$20\thinspace mm$\times$2\thinspace mm) and its associated electronics board with preamp, LNA and shaper. That actual componentes of the LNA are in the opposite side of the board to avoid electrical interference. The thick wire is the high-voltage supply to the detector (the blue wire is a test lead). {\it Bottom:\/} a computer rendering of an exploded view of the detector system. The upper plate is a structural Al frame in which a 0.016mm-thick aluminum foil will be stretched to provide light and electrical shielding to the detectors. The bottom brass piece provides structure and casing for the detectors and PCBs.}
\label{fig:detector+placa}
\end{figure}

Each detector is connected directly to a printed circuit board (PCB) that carries the front-end analog electronics with a preamp, a low-noise amplifier (LNA) and a shaper. The PCBs were designed in our lab and the first 10 were built in house. A contractor will replicate them for the whole detector plane. Figure \ref{fig:detector+placa} shows a photo of one detector and its associated board as well as an exploded view of the detector system. The detectors are mounted at 45$^{\circ}$\ with respect to the boards in order to minimize the spacing between adjacent detectors. 

The readout circuit of each detector has an applied bias voltage of $\sim -220$\thinspace V, a 100\thinspace M$\Omega$ bias resistor and a 150\thinspace nF coupling capacitor. The charge released by an X-ray interaction in each detector will feed an AC-coupled charge-sensitive preamplifier that creates an output pulse of typically $\lesssim 1$\thinspace mV. The pulses are then amplified by $\sim$65\thinspace dB by the LNA and formatted by a shaper. A Wilkinson-type 8-bit analog-to-digital converter (ADC) converts the pulse amplitudes to time and then measures this time. The digitised times are proportional to the energy deposited on the detector. This conversion technique is highly linear and precise; we find it is very suitable for CZT detectors with not very high count rates. 

\begin{figure}[!ht]
\centering
\includegraphics[width=0.9\hsize,angle=0]{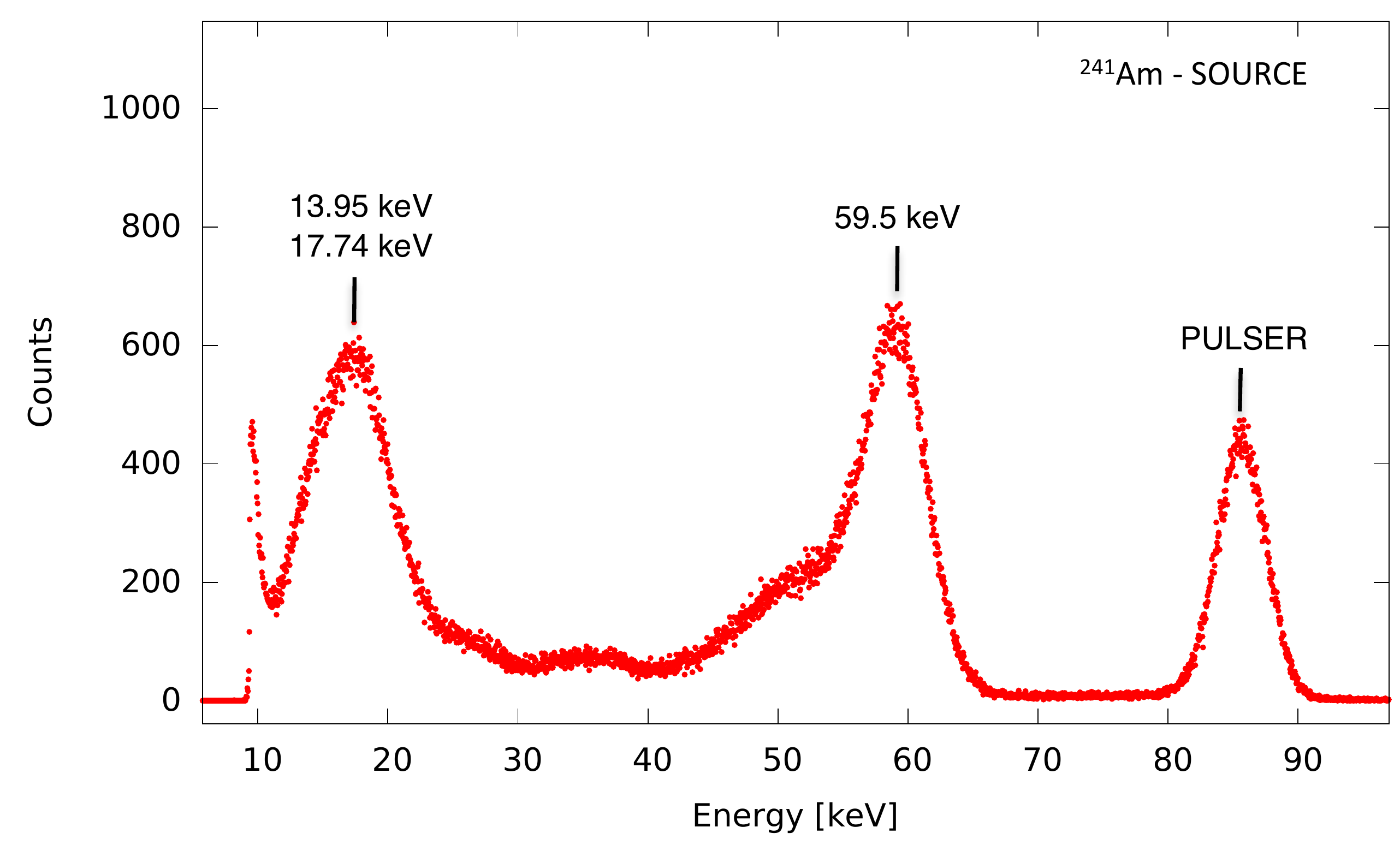}
\caption{Energy spectrum of a radioactive $^{241}$Am source for one of the CZT detectors. The peak on the right ($\sim 85$ keV) is produced by a pulser and has a FWHM of 4.5\thinspace keV. The 59.5\thinspace keV peak has a resolution of 6.6\thinspace keV FWHM ($\Delta E/E = 11$\thinspace \%).}
\label{fig:Am_spectrum}
\end{figure}

Figure \ref{fig:Am_spectrum} shows an energy spectrum of a radioactive $^{241}$Am source for one detector in the lab. The energy resolution at the 59.5\thinspace keV line is 6.6\thinspace keV\thinspace FWHM (``Full Width at Half Maximum''), corresponding to $\Delta E/E=11$\thinspace\%. Also shown is a peak from a pulse generator that we have used to measure the pulse height spread due to electronic noise. The equivalent electronic noise resolution is 4.5\thinspace keV FWHM. Since the electronic noise and the Poissonian fluctuations due to the number of charge carriers (electron-hole pairs) within the material must add in quadrature, the intrinsic, purely statistical resolution is 4.8\thinspace keV FWHM (8\%). The red wing on the 60\thinspace keV peak is due to incomplete charge collection within the CZT. The spectrum also shows a blend of Cd and Te escape peaks between the radioactive lines.

\subsubsection{The scientific data acquisition subsystem}

The pulses generated by X-ray interactions in the detectors are processed individually by a data acquisition scheme that includes the front-end analog electronics, described above, and a digital electronics for processing and formatting. The detector pulses are digitised, tagged with position and time, formatted, and then sent to the on-board data-handling computer for storage and transmission to the ground.

The detector signals of each row of 13 detectors in the detector plane array feed into a box with the {\it conversion electronics\/} (CE) with a Complex Programmable Logic Device (CPLD) and microcontrollers. Each one of the 13 CEs acquires data coming from the 13 detector PCBs on that particular row and feeds the pulses into 13 8-bit ADCs (mentioned above), one for each detector, so the energy of each event is determined in 256 channels. The CEs identify the specific detector that is hit on its associated row. Each event will be time-tagged unambiguously with 5\thinspace $\mu$s precision by an internal clock whose accuracy is checked each second by a GPS receiver. Each CE box receives PPS signals from the GPS.

The signals processed by the CEs will be read by a {\it multiplexing electronics\/} (MUX) box at a rate of 1\thinspace Hz. The MUX is thus responsible for the identification of the detector in which the event occurred (the event {\it position}), the energy and the relative time of the event. For each acquisition cycle (1 second), the MUX will attach to the data package the absolute time information coming from the GPS, so each event will have its instant of occurrence completely determined within 5\thinspace $\mu$s.  

The MUX will send to the on-board data handling computer (described in section \ref{sec:obdh}) complete datasets containing all data originated in the 13 CEs at a maximum rate of 115,200 bps. In this way, the time, position and energy of each event will be stored individually so that the position and energy distribution of events for any desired integration time can be determined. With an estimated count rate of $\sim$ 55\thinspace counts/s for the entire detector plane (see section \ref{sec:bkg}), the dead time will be negligible even for very strong sources at levels of several Crabs. 

\begin{figure}[!ht]
\centering
\includegraphics[width=0.9\hsize,angle=0]{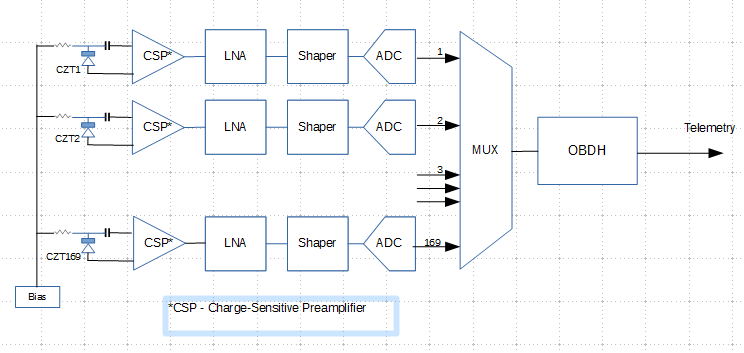}
\caption{Simplified block diagram of the detector data acquisition subsystem. The bias voltage is $\sim -200$\thinspace V, the bias resistor is of 100\thinspace M$\Omega$ and the coupling capacitance is of 150\thinspace nF. LNA: low-noise amplifier; ADC: analog-to-digital converter; MUX: multiplexing electronics; OBDH: on-board data handing computer.}
\label{fig:diagrama_Pepino}
\end{figure}

Figure \ref{fig:diagrama_Pepino} shows a simplified block diagram of the detector data acquisition subsystem.

\subsubsection{The collimator}

A collimator will be mounted in front of the detector plane in order to define the field-of-view (FOV), minimize background and provide a uniform illumination fraction of the detectors at any incidence angle. In this way, the entire FOV, all the way to zero intensity, will be fully coded by the mask, ensuring that the experiment does not have any {partially coded field-of-view (PCFOV). Other coded-mask systems do not use collimators, only shielding materials on the sides of the cameras. In these systems, sources in the PCFOV may create severe difficulties in the reconstruction process by casting incomplete shadowgrams of the mask pattern onto the detector plane. In addition, ambiguities in source positions are introduced. We have decided to use a collimator to avoid theses issues and reduce background. The price to be paid is the reduction of the sensitivity with increasing off-axis incidence angle, according to the collimator angular response}.

The collimator cell centers will be separated by 20\thinspace mm, in such a way that the individual CZT detectors will each be in the bottom of a collimator cell. The collimator blades are 81-mm high and have lead cores (0.5\thinspace mm) and copper plates (0.5\thinspace mm) on each side to absorb the radiation impinging at high angles and also provide a graded shield for lead X-ray fluorescence.

The collimator is surrounded by a graded shield made of lead (1.5\thinspace mm) in the outside and copper (0.5\thinspace mm) in the inside to suppress radiation from outside the field-of-view. The shielding walls have the same height as the collimator blades (81\thinspace mm) and are separated from the last blades by 2\thinspace mm, so that the environment that the last detectors (edges and corners) are subjected to are very similar to the more central ones. This is important to minimize inhomogeneities in the count distribution over the detector plane.

\subsubsection{The coded mask}
\label{sec:codedmask}

\begin{figure}[!ht]
\centering
\includegraphics[width=0.6\hsize,angle=0]{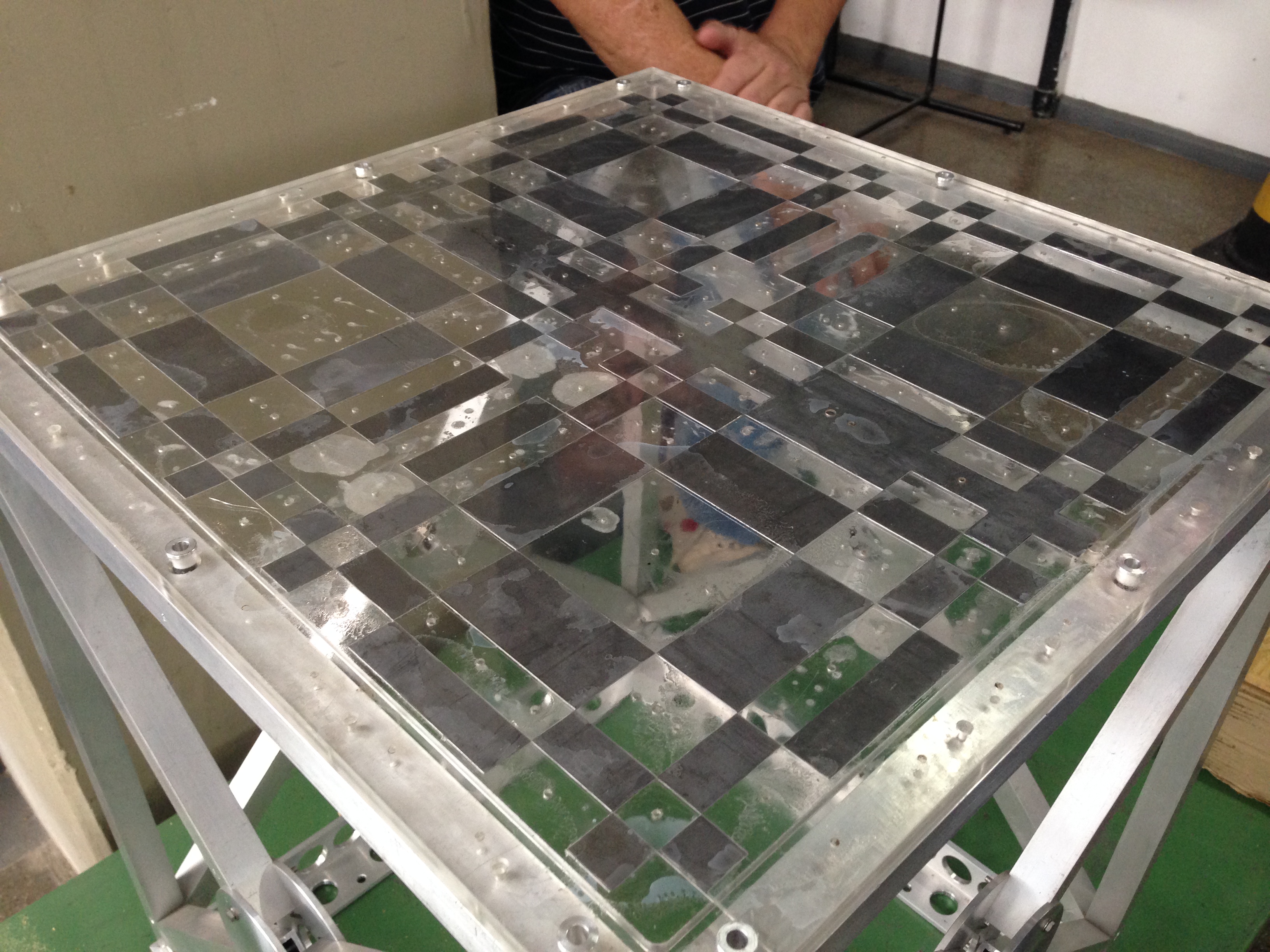}
\caption{Photo of the protoMIRAX coded mask. The basic elements are 1mm-thick 20\thinspace mm$\times$20\thinspace mm lead squares. The basic pattern is a  13$\times$13 MURA repeated four times (2$\times$2) minus 1 row and 1 column to provide unique shadowgrams for any incident direction (skybin). The total size is 500\thinspace mm$\times$500\thinspace mm. The lead pieces are glued to an X-ray transparent, 1mm-thick acrylic substrate both above and below the mask.}
\label{fig:coded mask}
\end{figure}

A coded mask, shown in Figure \ref{fig:coded mask}, will be mounted at a distance of 650\thinspace mm from the detector plane. The mask has the function of providing a spatial encoding of the X-ray incoming fluxes \citep[see][for an introduction of the concept]{1968ApJ...153L.101D}. With the knowledge of the mask pattern and the number of X rays detected by each CZT detector for a given integration time, one can reconstruct an image of the observed field-of-view by suitable deconvolution or cross-correlation algorithms \citep{1978ApOpt..17..337F}. The mask is made out of 1mm-thick lead sheets and the basic cells are 2\thinspace mm$\times$2\thinspace mm squares. The pattern of closed and open elements are a cyclic extension (2$\times$2) of a $13 \times 13$ Modified Uniformly Redundant Array (MURA) pattern \citep{1989ApOpt..28.4344G} with 20mm-side square basic cells. Of the 169 elements of the pattern, 84 are open and 85 are closed, giving an open fraction of 0.497. In order not to have complete repetitions of the pattern, which would introduce ambiguities in source positions in the sky, we remove one row and one column from the total 2$\times$2 array of basic patterns, so that the final dimensions of the mask are $25 \times 25$ elements or 500\thinspace mm $\times$ 500\thinspace mm.

The MURA patterns belong to a class of aperture patterns that allow images to be produced with no intrinsic noise, which means that the autocorrelation function of the mask pattern (actually the cross-correlation between the mask pattern and the decoding function, which is almost exactly equal to the mask pattern, with the exception of one single element) is a delta function. For a point source, the reconstructed image is ``perfect" in the sense that it produces a peak in an otherwise completely flat image (see  \citet{1989ApOpt..28.4344G} and \citet{2002RScI...73.3619B} for details about imaging with MURAs). With this configuration, the protoMIRAX X-ray camera will have a total field-of-view of $20^{\circ}.8 \times 20^{\circ}.8$ totally coded by the mask and a geometrical angular resolution (one {\it skybin\/}) of $1^{\circ} 43'$. {Due to the fact that there are 10-mm gaps between adjacent detectors}, the fraction of the FOV with maximum sensitivity (total detector area illuminated) will be $4^{\circ} 05' \times  4^{\circ} 05'$. {In other words, the collimator blades will only start shadowing the detectors for incident directions outside this angular range, centered on the camera axis.} The source location accuracy \citep[see][]{1987SSRv...45..349C} will depend on the incident angle and will be limited by the precision of the pointing system. In the best case, it will be around $10'$  for a 10-$\sigma$ source in the center of the field of view. 

\subsubsection{The passive shield}

\begin{figure}[!ht]
\centering
\includegraphics[width=0.45\hsize,height=6cm,angle=0]{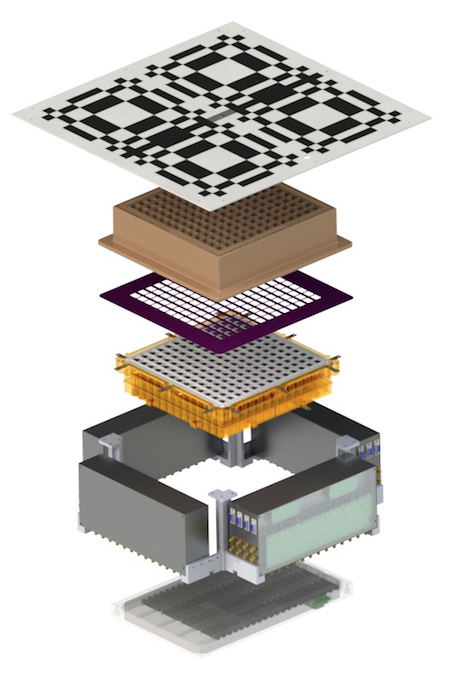}
\hspace{0.2cm}
\includegraphics[width=0.5\hsize,angle=0]{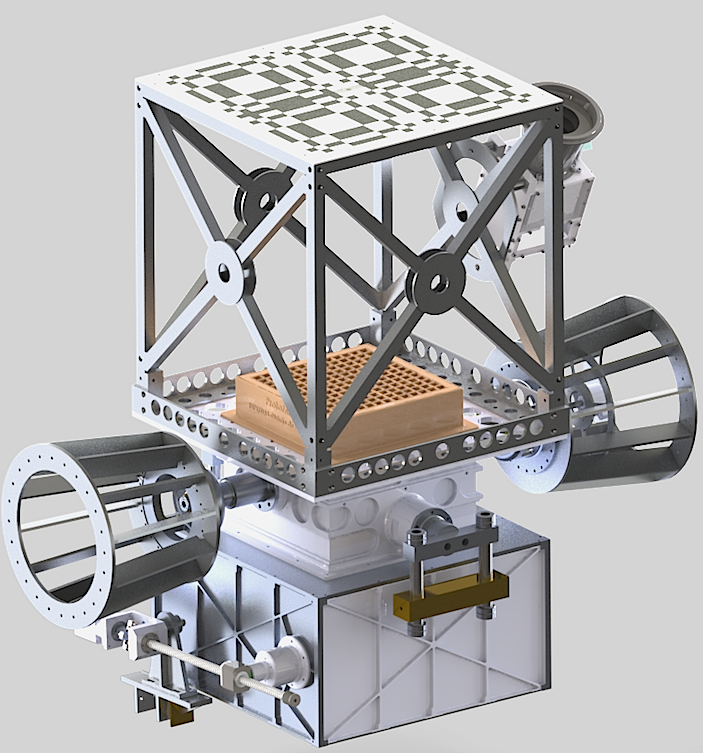}
\caption{Computer rendering of the protoMIRAX X-ray Camera. {\it Left:\/} Exploded diagram of the XRC, showing, from top to bottom, the coded mask, collimator, aluminum foil support plate, detector system, detector electronics and power supply boxes, and battery pack. {\it Right:\/} The assembled XRC with passive shielding and structural parts, including the elevation axis support. Also shown are one of protoMIRAX's  two star cameras (top right) and the elevation driving mechanism (bottom left) based on a ball screw.}
\label{fig:exploded_CRX}
\end{figure}

The detector array and associated electronics, including the detector boards, power supplies and batteries, will be  surrounded at the sides and bottom by a gradual passive shield of lead (1.5\thinspace mm) and copper (0.5\thinspace mm). This shield will be responsible for absorbing most of the radiation coming from directions other than those defined by the collimator. Figure \ref{fig:exploded_CRX} shows computer diagrams with the main parts of the XRC, including the aluminum support structures.

\subsection{The Balloon Gondola}
\label{sec:gondola}

\begin{figure}[!ht]
\centering
\includegraphics[width=0.8\hsize,angle=0]{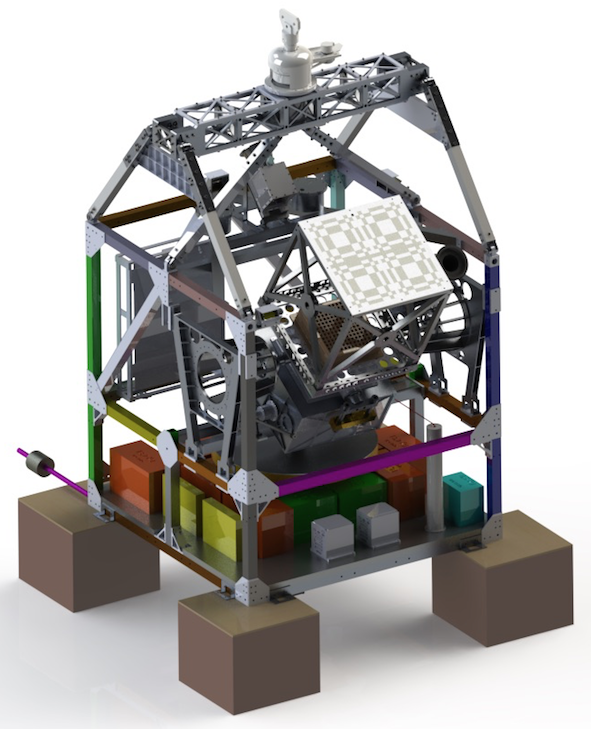}
\caption{Computer rendering of the protoMIRAX balloon gondola, showing the X-ray camera in the front face. The coloured boxes at the bottom of the frame are battery packs and electronics racks. The flat box in the back is the TM\&TC subsystem. The reaction wheel can be seen below the XRC. One can also see the two star cameras: one is fixed at the XRC and the other (behind the coded mask) is fixed in the gondola frame. At the top it is shown the decoupling mechanism that connects to the balloon cables. The four brown boxes in the bottom are crash pads. The gondola frame dimensions are 1.4\thinspace m $\times$ 1.4\thinspace m $\times$ 2\thinspace m.}
\label{fig:gondola}
\end{figure}

The protoMIRAX balloon gondola will house the X-ray camera and the various subsystems of the space segment, including the On-Board Data Handling Subsystem (OBDH), the Attitude Control and Pointing Subsystem (ACS), the Telemetry and Command Subsystem (TM\&TC), and the Power Supply Subsystem (PSS). Also, there will be a main GPS receiver that will provide universal time to all subsystems and two star cameras that will be part of the control loop. A computer rendering of the gondola is shown in Figure \ref{fig:gondola}. 

The gondola will use a power supply system based on packages of Li-ion batteries. A telemetry system operating in the L-band (1.5\thinspace GHz) will provide data and housekeeping links to the ground. A command system will provide pointing operations during the balloon flight. The science data will be stored on-board on computer flash memories and will also be entirely transmitted to the ground.

\subsubsection{The On-Board Data Handling Subsystem}
\label{sec:obdh}

The On-Board Data Handling Subsystem (OBDH) is responsible for acquiring, formatting and transmitting all data that come from the various subsystems of protoMIRAX's space segment to the ground station (GS). The OBDH is also responsible for receiving and retransmitting, when necessary, the various commands sent by the GS to the gondola. For each X-ray photon detected by the XRC, a 6-Byte packet is created 
encasing the time stamp, $x$-$y$ position (detector that was hit) and energy (pulse height) of the event that were formatted and sent by the MUX. These event packets are then sent every 1 second to the OBDH through a 115.2 kbps RS-422 unidirectional serial communication line. In addition, another 115.2 kbps RS-422 serial interface allows OBDH to send specific commands to the XRC and receive XRC's housekeeping data.  

We devised a cost-effective and robust solution for the experiment timing in which a single Global Positioning System (GPSDXA) unit sends PPS (1 pulse per second) and data to the various subsystems of the experiment: OBDH, XRC, ACS, and the Autonomous Star Trackers (AST). The OBDH attaches the UT data from the GPSDXA to the scientific data files at every second, so absolute timing information at the resolution of 5\thinspace $\mu$s can be retrieved for every event.

The OBDH also communicates with the GS and the ACS. A 500\thinspace kbps synchronous communication channel connects the OBDH with the Flight Control and Telecommunications Subsystem (FCTS). This channel is exclusively dedicated to transmit scientific data to ground. FCTS also communicates with the GS via a double transmitter/receiver system operating in L-band. In order to transmit all housekeeping data generated by the several subsystems of the space segment, OBDH uses a serial RS-232 channel operating at 115.2 kbps. Ground station commands are received by OBDH via FCTS by means of a 9.6 kbps RS-232 serial channel. In addition, all ACS data are sent to OBDH prior to be sent to the GS via another 115.2 kbps RS-422 serial interface standard.

The main hardware unit of OBDH is the Payload Data Handling Computer (PDCpM), which is a PC/104 ultra low power AMD Geode LX computer. It operates on 333 MHz, has a 128-MByte SDRAM, and has all the necessary interfaces previously mentioned in addition to analog-digital/digital-analog converters. The PDCpM went through thermal cycling and thermal vacuum testing with temperatures ranging from $-40$$^{\circ}$ C to $+85$$^{\circ}$ C and pressure as low as 1.3 mbar. 

The software embedded into the computer (SWPDCpM) has been developed in C language over the real-time operating system RTEMS.  The architecture of the SWPDCpM is composed of 3 layers where the bottom layer is the basic software, the intermediate layer is the flight software library in which there are basic services and reusable components, and the top layer is the application software, the main part of SWPDCpM.

\begin{figure}[!ht]
\centering
\vspace{-1cm}
\includegraphics[width=0.9\hsize, angle=0]{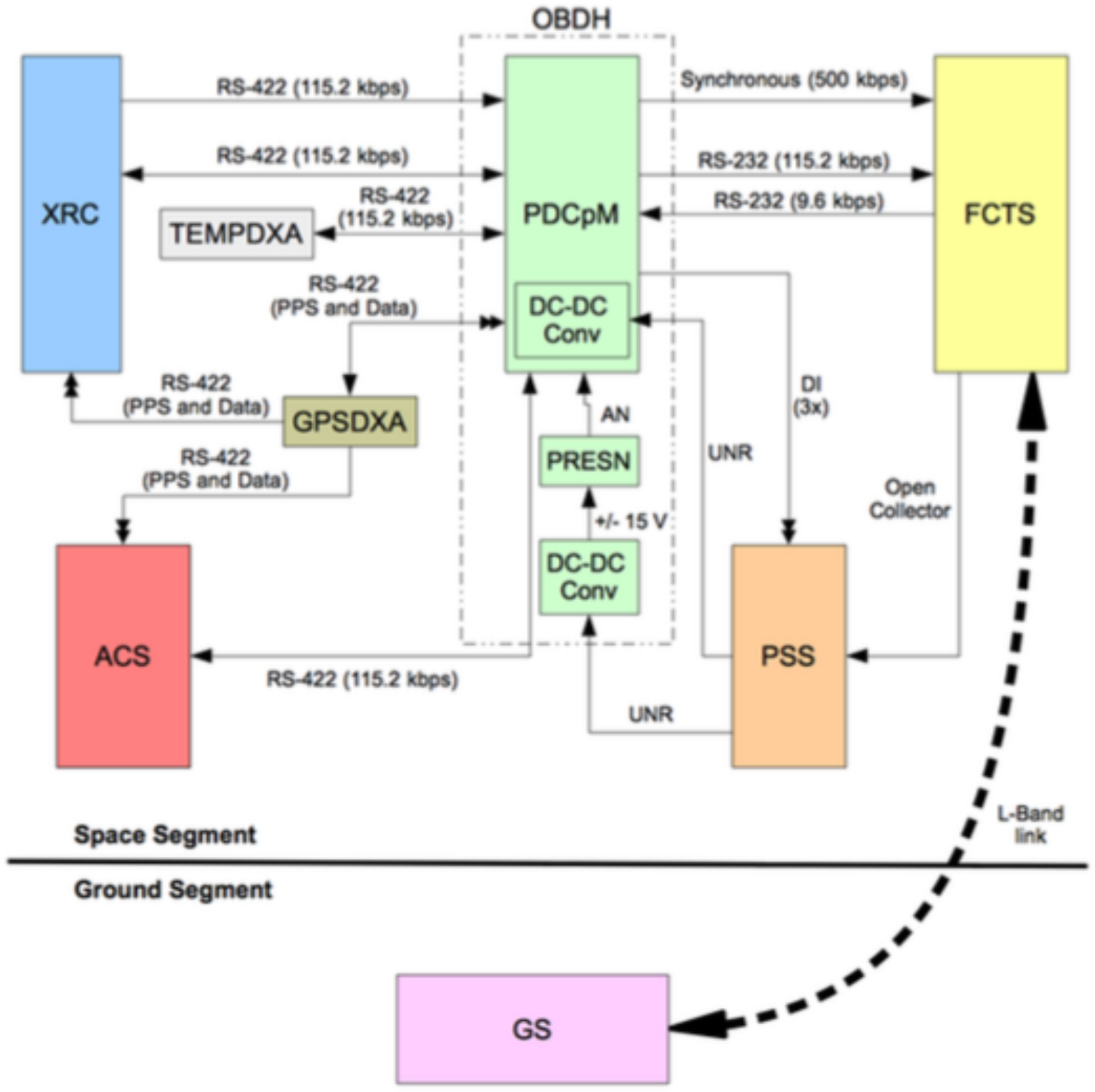}
\vspace{-2cm}
\caption{Simplified physical architecture of the protoMIRAX balloon experiment. XRC: X-ray camera; ACS: Attitude Control System; TEMPDXA: Temperature monitoring equipment; GPSDXA: GPS unit; OBDH: On-Board Data Handling Subsystem; PDCpM: Payload Data Handling Computer; DC-DC Conv: DC-DC Converter; PRESN: Pressure Sensor; PSS: Power Supply Subsystem; FCTS: Flight Control and Telecommunications Subsystem.}
\label{fig:block}
\end{figure}

Figure \ref{fig:block} shows a simplified physical architecture of the protoMIRAX experiment with its main subsystems.

\subsubsection{The Attitude Control and Pointing Subsystem}
\label{sec:sca}

The X-ray camera will be placed in an alt-azimuth mount in the balloon gondola, allowing pointing to different targets in the sky and tracking them continuously down to elevation angles of 30$^{\circ}$. A ball screw mechanism, with a step motor, will allow smooth and secure motion in elevation, and the azimuth motion will be provided by a system that rotates the entire gondola with respect to the geomagnetic field and the positions of the sun and the stars in the sky. The attitude control system will employ several sensors and actuators. The sensors include an electronic compass, an accelerometer, two star sensors, a sun tracker and both azimuth and elevation angular encoders. The actuators include a reaction wheel for angular momentum exchange and dump, and motors that will drive the motions in both elevation and azimuth. The gondola also has a mechanism at its top for decoupling its rotation with respect to the balloon itself and a motor for desaturating the reaction wheel.  The overall pointing accuracy of the control system is $\lesssim 0.5^{\circ}$, depending on the elevation angle; high elevations are in general more difficult to control since small changes in the pointing direction near the zenith leads to large azimuthal and roll angle displacements. Also, the accuracy will depend on whether the flight is at daytime or nighttime. At nighttime, the star cameras will provide a more accurate control. 

The two autonomous star sensors of protoMIRAX are being developed at INPE. They are wide-field (25$^{\circ}$$\times$25$^{\circ}$) optical (400$-$800\thinspace nm) cameras with a 35\thinspace mm focal length. The imaging sensor is a CMOS APS (active pixel sensor) with 1024$\times$1024 pixels and quantum efficiency up to 25\%. An especially designed algorithm for star pattern recognition provides attitude knowledge precision better than 10\thinspace arcseconds (1\thinspace $\sigma$) around the cross-foresight axes (yaw and pitch) and better than 60\thinspace arc seconds (1\thinspace $\sigma$) around the foresight axis (roll angle). One of the star sensors is fixed in the XRC mechanical frame so that it enables precise pointing of the XRC telescope. Its optical axis is tilted 24$^{\circ}$\ with respect to the X-ray axis, toward lower elevations, so as to avoid having the balloon in the field of view for high elevation observations. Even with this misalignment, there will be occasions when the primary star tracker will be obstructed by the balloon. In these cases the XRC attitude will be determined by the secondary star tracker mounted in the gondola, albeit with a somewhat reduced accuracy due to uncertainties from the angular encoders used in the motors that control the XRC mechanical frame elevation.
The star trackers are also capable of acquiring images, which will be used for evaluation of the star tracker performance during the flight.  

\subsection{The Ground Subsystem}

The operations of the protoMIRAX telescope during the balloon flights will make use of the ground facilities already available at INPE. The reuse strategy focused on two subsystems: (i) the balloon flight control (OPS) and (ii) the experiment operation itself, which will be supported by the SATellite Control System (SATCS). SACTS is a software-based architecture developed at INPE for satellite commanding and monitoring. Considering that protoMIRAX is a pathfinder for the MIRAX satellite mission, a ground infrastructure compatible with INPE's satellite operation approach would be useful and highly recommended to control and monitor the experiment during the balloon campaigns. 
 
The OPS required investments on information technology. The available balloon ground station needed little hardware maintenance, but significant efforts on software updating were necessary. The original standalone computer system dedicated to run the flight control software was modernized both in hardware and software by the new system named OPS/ES. In addition, a new server computer, properly configured for Ethernet connections, has extended the existing ground station facilities with a network switch, serial converters and a new software named OPS/Server in order to support the available uplink and downlink channels being mapped to TCP/IP gateways. Those communication improvements were necessary to support the interoperability between the balloon ground station, SATCS and protoMIRAX data center. 

\begin{figure}[!ht]
\centering
\includegraphics[width=0.9\hsize]{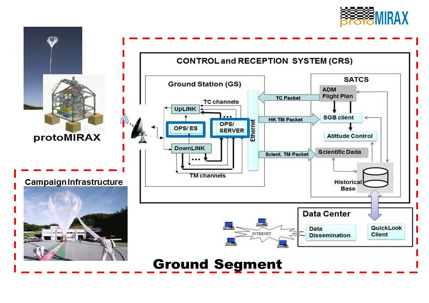}
\caption{Physical architecture of the protoMIRAX ground subsystem.}
\label{fig:solo}
\end{figure}

Concerning SATCS, some effort in software customization was necessary because SATCS architecture complies with particular operational requirements on different missions by using several customized object-oriented software elements and frameworks. The diagram in Figure \ref{fig:solo} shows the ground solution designed for the Ground Segment of the protoMIRAX experiment.

\section{Effective Area and Sensitivity}
\label{sec:bkg}

The effective area of a balloon experiment for the observation of cosmic sources is a function of energy ($E$), atmospheric depth and zenith angle:
\begin{equation}
A_{\rm eff} (E) = A_{\rm geo} \; \epsilon\; e^{-(\mu/{\!\rho})_{\!A} \, x \sec z}
\end{equation}
where $A_{\rm geo}$ is the geometrical detection area, $\epsilon \equiv 1 - \exp({-\mu_{\!D}\, l})$ is the detector efficiency, $\mu_{\!D} (E)$ is the detector's attenuation coefficient in cm$^{-1}$, $l$ is the detector thickness in cm, $(\mu/{\!\rho})_{\!A} (E)$ is the attenuation coefficient of the atmosphere in cm$^2$/g, $x$ is the atmospheric depth in g cm$^{-2}$ and $z$ is the zenith angle.  

\begin{figure}[!ht]
\centering
\includegraphics[width=0.9\hsize,angle=0]{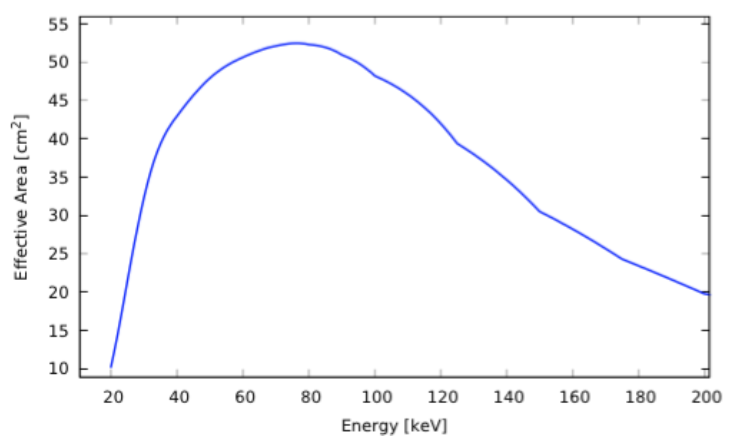}
\caption{Effective area of protoMIRAX.}
\label{fig:aeff}
\end{figure}

For the expected observations of protoMIRAX at 2.7 g cm$^{-2}$, considering a point source at the zenith, the effective area is shown in Figure \ref{fig:aeff}, with a maximum value of 52 cm$^2$ at $\sim$80\thinspace keV. It is important to note that the geometrical area for a point source at infinity is the total detector area of 169\thinspace cm$^2$ multiplied by the open fraction of the mask, which is 0.497 (see section\ref{sec:codedmask}).

The sensitivity of the experiment can be calculated by estimating the background at the altitudes expected for the balloon flight. This was produced by performing Monte Carlo simulations using the software package GEANT4 \citep{2003NIMPA.506..250A}, a well-known suite of routines that performs detailed calculations of gamma-ray and particle interactions on all materials. Given the experiment mass model and the environmental particle and photon fields, the code provides the detector spectral response, both prompt and delayed. 

Based on detailed GEANT4 calculations we have carried out {(\citet{2015Penacchionietal}; Castro et al., in preparation)}, the estimated background count rate for the entire detector plane, in the 30 to 200\thinspace keV energy range, is of $\sim$ 55\thinspace counts/s. {In these simulations, we have considered both the anisotropic atmospheric X and $\gamma$-ray spectra and the cosmic diffuse contribution coming from the telescope aperture. We have also considered primary and secondary protons, electrons and neutrons. As shown by \citet{2015Penacchionietal}, the photon contribution is dominant up to 90 keV, where the contribution due to protons becomes almost equally important. Neutrons are important below 35 keV and the electron contribution is negligible.}

\begin{figure}[!ht]
\centering
\includegraphics[width=0.8\hsize,height=6cm,angle=0]{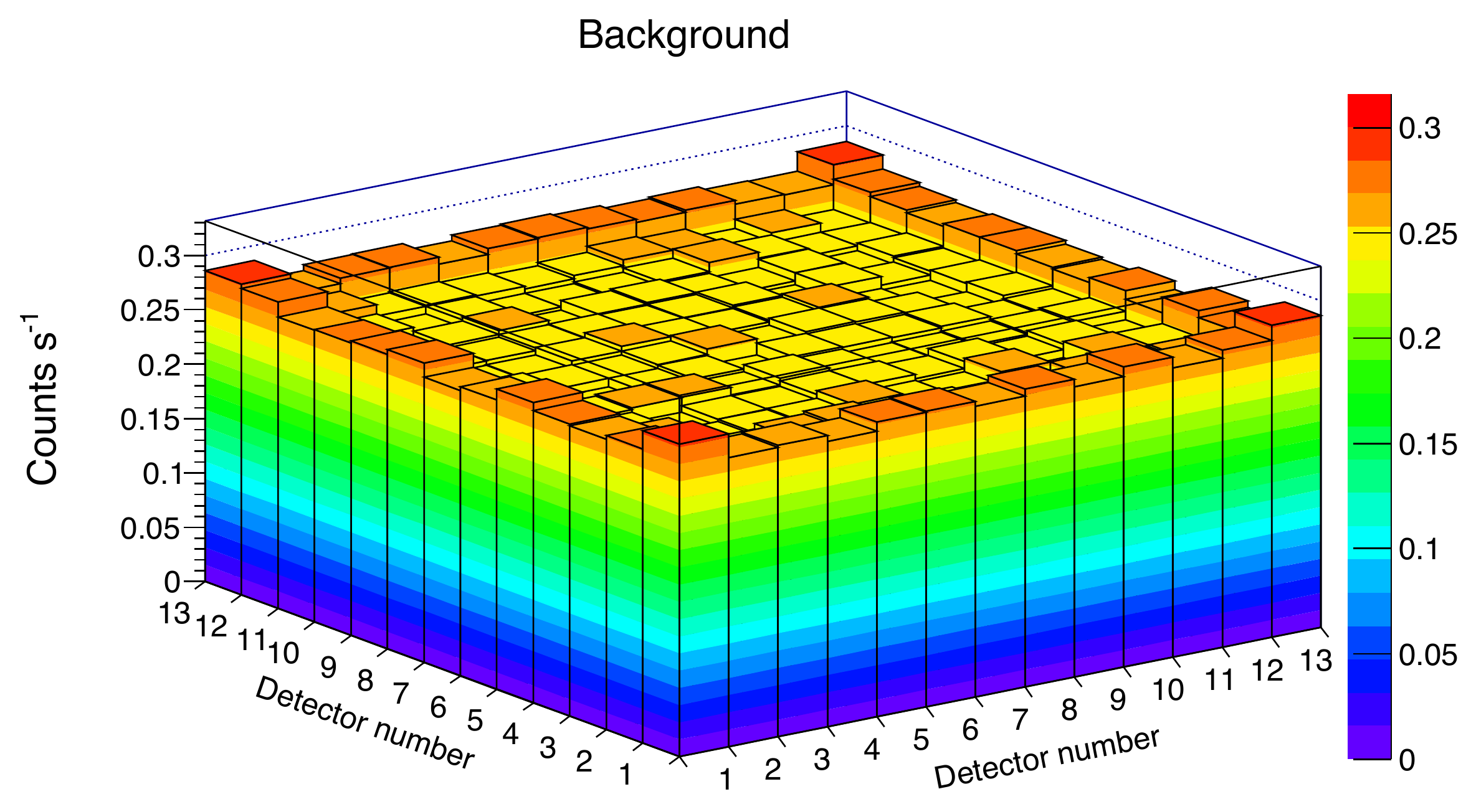}
\caption{Background count distribution over the detection plane in the detector energy range 30 - 200\thinspace keV, for an observation time of 4\thinspace h at balloon altitudes.}
\label{fig:shadowgram_bkg}
\end{figure}

In Figure \ref{fig:shadowgram_bkg} we show a map of the simulated background counts over the detector plane for an integration of 4 hours at 2.7 g cm$^{-2}$ for the entire energy range 30-200 keV. The increased count rates at the edges and corners of the detector plane {are most likely} due to the closeness of these detectors with the shielding materials and support structures, which produce secondary radiation by scattering. {Another factor could be due to the extra shielding produced by the collimator blades to the detector in the central areas of the detector plane.} 

Taking the expected background levels at balloon altitudes as a function of energy, one can calculate the sensitivity of the experiment in terms of the minimum detectable flux at a particular statistical significance. As shown by \citet{1989ApOpt..28.4344G}, the signal-to-noise-ratio (SNR) of a coded mask instrument that uses URAs or MURAs, for a single point source, is given by 
\begin{equation}
SNR = \frac{N_S}{\sqrt{2 N_S + N_B}} ,
\end{equation}
where $N_S$ is the net source counts and $N_B$ is the background number of counts for a given integration time in a particular energy range. 
Solving for $N_S$, one gets
\begin{equation}
N_S = SNR\;  \left( SNR + \sqrt{SNR^2 + N_B} \right).
\end{equation}
If we are interested in a {\it minimum detectable flux\/}, the SNR will be a small number ($\lesssim 10$), and it is reasonable to assume that $SNR^2 \ll N_B$ for sufficiently long integration times. Therefore, $N_S \approx SNR\, \sqrt{N_B}$.
 
Now, if $S$ is the source flux in photons cm$^{-2}$s$^{-1}$keV$^{-1}$, then $N_S = S\, A_{\rm eff} \, T\, \Delta E$, where  $A_{\rm eff}$ is the effective area in cm$^2$, $T$ is the integration time in seconds and $\Delta E$ is the energy interval under consideration (in keV). Similarly,  $N_B = B\, A_{\rm geo} \, T\, \Delta E$, where $B$ is the background level in the more suitable units of counts cm$^{-2}$s$^{-1}$ and $A_{\rm geo}$ is the geometrical area of the whole detector plane. 

\begin{figure}[!ht]
\centering
\includegraphics[width=0.95\hsize,angle=0]{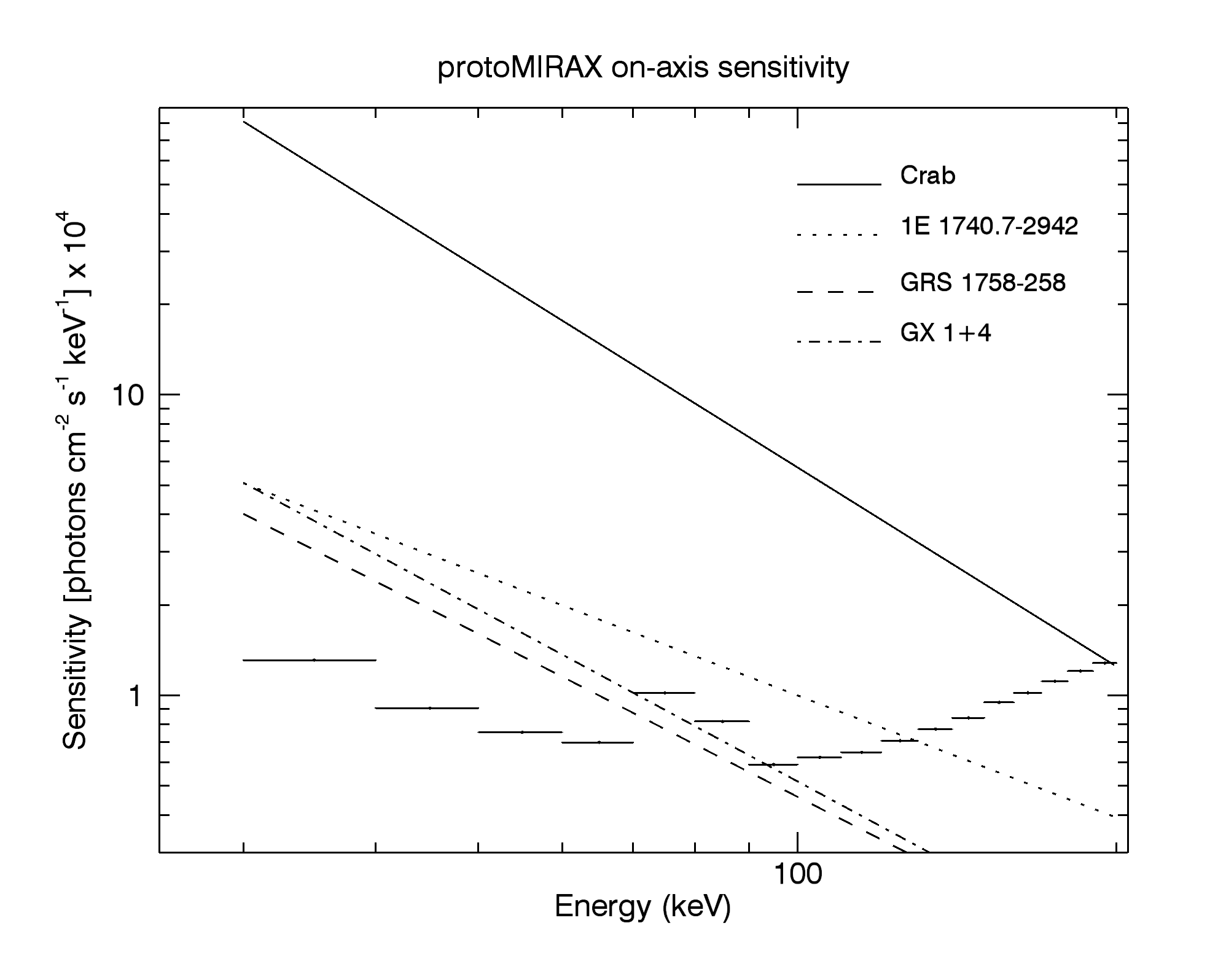}
\caption{{On-axis} sensitivity curve for protoMIRAX. The horizontal bars are the minimum detectable fluxes at a level of 3$\sigma$ at an atmospheric depth of 2.7\thinspace g cm$^{-2}$ and a zenith angle of 30$^{\circ}$. The integration time is 8\thinspace hours. Also shown are the spectra of 4 X-ray sources that will be observed in the first balloon flight. The source spectra were taken from \citet{2004ESASP.552..815S} (Crab), \citet{1995AdSpR..15..115G} (1E\thinspace $1740.7-2942$), \citet{1991SvAL...17...50S} (GRS\thinspace $1758-258$) and \citet{1991AdSpR..11...35D} (GX\thinspace $1+4$).}
\label{fig:sensitivity}
\end{figure}

The minimum source flux that will be detectable at a level of $N_{\sigma} \equiv SNR$ will then be
\begin{equation}
F_{\rm min} = \frac{N_{\sigma}}{A_{\rm eff}\, \Delta E}\; \sqrt{\frac{B \, A_{\rm geo}}{T}} \;\; {\rm photons\; cm^2 s^{-1} keV^{-1}}.
\end{equation}
According to this expression, the $3$-$\sigma$ sensitivity (minimum detectable flux) for the whole energy band is found to be approximately $1.9 \times 10^{-5}$ counts cm$^{-2}$ s$^{-1}$ from 30 to 200 keV, considering 8\thinspace h of integration at 2.7 g cm$^{-2}$ and a zenith angle of 30$^{\circ}$. This allows protoMIRAX to make detailed observations of the Crab (even at low elevation angles) and also detect a few X-ray binaries in states of high hard X-ray emission. In Figure \ref{fig:sensitivity} we show a 3-$\sigma$ sensitivity curve for protoMIRAX for an integration time of 8 hours at 2.7 g cm$^{-2}$, considering a point source at a zenith angle of 30$^{\circ}$. The increased minimum flux values between 70 and 90\thinspace keV are due to somewhat strong lead fluorescent lines (reported by \citet{2015Penacchionietal}) in the estimated background spectrum. We are in the process of studying different shielding configurations that will very likely minimize this problem. 

\section{Image Simulation}
\label{sec:image}

\begin{figure}[!ht]
\centering
\includegraphics[width=0.7\hsize, height=6cm, angle=0]{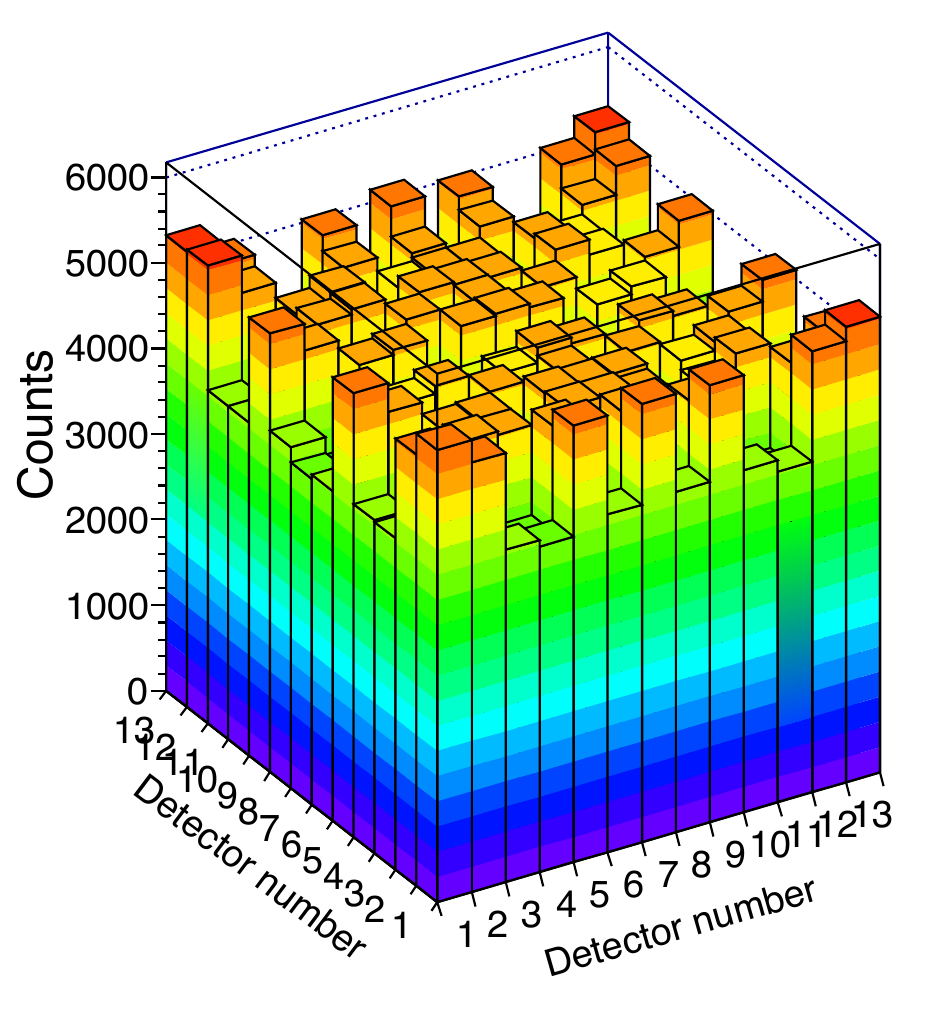}
\caption{A count distribution (shadowgram) on the detector plane of a simulated observation of the Crab, including the expected background, for an integration time of 4\thinspace h around the meridian passage at a latitude of 23$^{\circ}$\ S. The energy range is 30 to 200 keV.}
\label{fig:shadowgram_crab}
\end{figure}

In order to predict the performance of the protoMIRAX imaging system, we have simulated an observation of the Crab nebula during a meridian transit at our flight latitudes (23$^{\circ}$ S). To calculate the number of counts from the source, we have considered the zenith angle variation with time before and after the transit, which occurs at an elevation of 45$^{\circ}$. Based on the Crab hard X-ray spectrum taken from \citet{2004ESASP.552..815S}, we built with GEANT4 a simulated shadowgram of a Crab observation in the center of the FOV, which is shown in Figure \ref{fig:shadowgram_crab}.

\begin{figure}[!ht]
\centering
\includegraphics[width=0.8\hsize,height=0.5\hsize,angle=0]{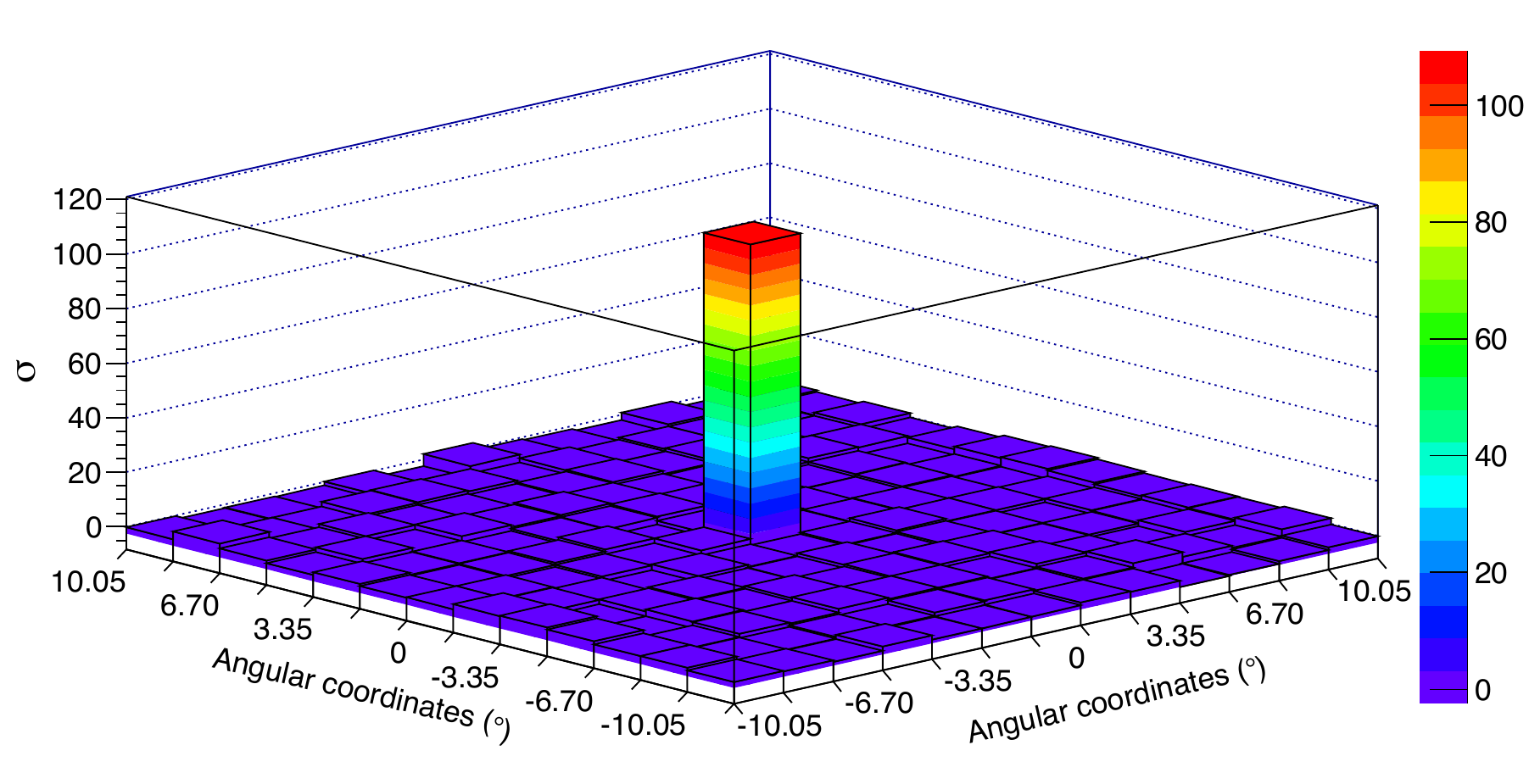}
\caption{Simulated image of the Crab Nebula region for an integration time of 4\thinspace h around the meridian transit at balloon altitudes (2.7\thinspace g cm$^{-2}$). The energy range is 30$-$200\thinspace keV. The signal-to-noise ratio of the detection is 109.}
\label{fig:Crabsim}
\end{figure}

The image covers the entire FOV of 21$^{\circ}$ $\times$ 21$^{\circ}$, and each sky bin corresponds to 1$^{\circ}43^{\prime}$ of the sky. The overall signal-to-noise ratio (SNR) of the image is 109\thinspace $\sigma$ \citep[see][for details]{2015Penacchionietal}.  This value corresponds to 91\% of the theoretical value given by expression (2) above. The loss in detection significance is mainly due to the non-uniformity of the background across the detector plane. In Figure \ref{fig:Crabsim} we show the reconstructed Crab image. It is interesting to note that the Crab will be in principle observable by protoMIRAX at a $5$-$\sigma$ level for an integration as short as 63\thinspace s, which makes it a very good calibration source for the experiment.

\section{Conclusion}
\label{sec:conclusion}

The protoMIRAX experiment is a balloon-borne hard X-ray telescope being developed as a scientific and technical pathfinder for the MIRAX satellite mission. {In the current configuration, under revision due to restrictions imposed by the Brazilian space program, MIRAX will consist in a set of wide-field coded-mask hard X-ray cameras (5$-$200 keV) with an angular resolution of $\sim$ 5$-$6 arcmin that will operate in scanning mode in a near-equatorial circular low-Earth orbit. MIRAX main goals include the characterization, with unprecedented depth and time coverage, of a large sample of variable and transient phenomena on accreting neutron stars and black holes, as well as Active Galactic Nuclei and GRBs.} 

Apart from serving as a testbed for several MIRAX subsystems, protoMIRAX will be capable of doing interesting science of its own. The hard X-ray camera, similar to the hard X-ray imager being developed for MIRAX, albeit with a much lower angular resolution and smaller detector area, has a resolution of $1^{\circ}43'$ with a fully-coded FOV of $21^{\circ} \times 21^{\circ}$. Detailed background calculations have been performed with the GEANT4 package, and the expected background at balloon altitudes at a latitude of $-23^{\circ}$ over the SAA region in Brazil provides a $3$-$\sigma$ sensitivity for the 30$-$200 keV range of $\sim1.9 \times 10^{-5}$ photons cm${^-2}$ s$^{-1}$ for an integration of 8\thinspace hours. In its first balloon flight, protoMIRAX will observe the Crab Nebula for calibration and imaging demonstration and also the GC region for scientific purposes. The experiment is in a very advanced state of development and we hope to carry out the first ballon flight in late 2015. 

\begin{acknowledgements} We thank FINEP, CNPq and FAPESP, Brazil, for financial support. We also thank Fernando G. Blanco, S\'ergio Admir\'abile, Luiz A. Reitano, Wendell P. da Silva, Leonardo Pinheiro, Fernando Orsatti, Paulino Scherer and the engineers from the company COMPSIS in S\~ao Jos\'e dos Campos\ for invaluable technical support. A.V.P.\ acknowledges the support by the international Cooperation Program CAPES-ICRANET financed by CAPES - Brazilian Federal Agency for Support and Evaluation of Graduate Education within the Ministry of Education of Brazil. M.C.A.\ and J.R.S.\ also acknowledge CAPES for Ph.D. and post-doctoral fellowships, respectively.
\end{acknowledgements}

 \bibliographystyle{aa} 
 \bibliography{paper_pMIRAX_AA_final}

\end{document}